\documentclass[twocolumn,aps,showpacs]{revtex4}
\usepackage{amssymb}
\usepackage{epsfig}

\newcommand{\be}{\begin{equation}}
\newcommand{\ee}{\end{equation}}
\newcommand{\beq}{\begin{eqnarray}}
\newcommand{\eeq}{\end{eqnarray}}
\newcommand{\Dirac}{\rlap{\hspace{-.6mm} \slash} D}
\newcommand{\nn}{\nonumber}

\def\ga{\gamma}
\def\wa{w^{(a)}}
\def\al{\alpha}                            
\def\fat{f^{(a)}(\tau)}
\def\zc{{z^\ast}}
\def\uc{{u^\ast}}

\def\re{{\Re\mbox{e}}}
\def\im{{\Im\mbox{m}}}

\begin{document}
\title{Microscopic correlation functions for the QCD Dirac operator 
with chemical potential}
\author{G. Akemann}
\affiliation{
Service de Physique Th\'eorique, CEA/Saclay, 
F-91191 Gif-sur-Yvette Cedex, France}


\begin{abstract}
A chiral random matrix model with complex eigenvalues is 
solved as an effective model for QCD with non-vanishing 
chemical potential. We derive new matrix model correlation functions which 
predict the local fluctuations of 
complex Dirac operator eigenvalues at zero virtuality. 
The parameter which measures the non-Hermiticity of the Dirac
matrix is identified with the chemical potential. 
In the phase with broken chiral symmetry all spectral correlations of the Dirac
eigenvalues are calculated as functions of quark flavors 
and chemical potential. The derivation uses the orthogonality of the Laguerre
polynomials in the complex plane. Explicit results are given for any 
finite matrix size $N$
as well in the large-$N$ limit for weak and strong non-Hermiticity. 
\end{abstract}

\maketitle   


Random Matrix Models (RMM) have been a useful tool in Theoretical Physics 
for a long time.
In many physical systems the local 
fluctuation properties for example of the energy levels are universal and 
can be successfully described by RMM, where we refer to \cite{GMW} for a 
review. Although in the generic situation the Hamiltonian and other physical 
observables are Hermitian thus having real eigenvalues, there 
exist also important cases where complex eigenvalues occur. 
As examples we mention localization in superconductors \cite{HN}, 
dissipation and scattering 
in Quantum Chaos \cite{GHS} or Quantum Chromodynamics (QCD) 
with chemical potential \cite{Steph}.

Much less is known so far for spectral correlations of complex eigenvalues
derived from RMM. Although the first results date back to 
Ginibre \cite{Gin} where the correlations for the complex Unitary Ensemble 
labeled by the Dyson index $\beta=2$ were 
calculated, progress has been slow. The correlation functions 
of the ensemble with real non-symmetric 
matrices ($\beta=1$) are still unknown. 
Results for quaternion matrices ($\beta=4$) were obtained 
more recently in \cite{Mehta,EK} and the 
inclusion of Dirac mass terms for $\beta=2$ in  
\cite{A01}. Furthermore, it has been realized by works of 
Fyodorov and collaborators \cite{FKS}
that different regimes of complex eigenvalues exist, the 
weak and strong non-Hermiticity limit. In the present work we wish to 
extend the knowledge about 
complex matrix models to the so-called chiral models. 

Chiral RMM of real eigenvalues have been introduced 
to describe the local fluctuation properties of the Dirac operator in QCD
at the origin \cite{SV}. 
The low energy spectrum of the QCD Dirac operator is a very sensitive tool
to study the phenomenon of chiral symmetry breaking \cite{JT}.
The predictions of the different chiral RMM ensembles have been very 
successful in describing the dependence on the gauge group and its 
representation, the number of quark flavors and masses and the topology 
\cite{lattRMT}.
In particular the topology dependence has been very useful in comparison
with new developments in lattice gauge theory, admitting to 
incorporate an exact chiral symmetry \cite{Nie}. 
By now the field theoretic origin of the RMM description has also been well 
understood \cite{DOTV}. 

On the other hand lattice simulations in the presence of a chemical 
potential $\mu$, which renders the Dirac eigenvalues complex, remain extremely 
difficult, as reviewed in \cite{lattmu}. 
Although recent progress has been made 
\cite{Fodor} along the phase transition line the general phase diagram 
remains unexplored for $\mu\neq 0$. 
It is in this context that analytical knowledge 
of microscopic correlation functions from 
complex chiral RMM could be very useful, in view of its success 
in predicting real eigenvalue correlations.

Chiral RMM including a chemical potential have been already studied in 
several works \cite{Steph,HJV,Halasz,MPW}. 
In \cite{Steph} the nature of the quenched limit has been 
analyzed in such a model, and the global density of complex eigenvalues 
together with  
its boundary have been calculated as a function of $\mu$. The phase diagram 
of QCD in the temperature density plane has also been predicted with such a 
model \cite{Halasz}. 
Recently complex Dirac eigenvalues calculated on the lattice 
have been confronted to a complex matrix model on the microscopic scale
given by the inverse volume in the bulk of the spectrum
\cite{MPW}. The nearest neighbor distribution 
along a given direction in the complex plane was considered and a transition 
from the Unitary to the Ginibre ensemble was observed at increasing $\mu$, 
ending in a Poisson distribution.

Our aim here is to provide more detailed information by
calculating all microscopic correlation functions in the complex plane, 
both in the limit of weak and strong non-Hermiticity. 
Our model gives a natural chiral extension of the ensemble treated in 
\cite{FKS,A01}. Although our main motivation is the application to QCD 
lattice calculations with $\mu\neq 0$ the new correlation functions 
we derive may find other applications as well, 
such as in the fractional Quantum Hall effect \cite{PdF}
or two-dimensional charged plasmas \cite{Janco}.
We restrict ourselves to work with a Gaussian chiral RMM.
In particular 
we will not touch the issue of universality here although we expect that in 
analogy to \cite{ADMN} the same correlation functions hold for a more general
weight function at large-$N$. 

We start our investigation by defining our model. The chiral random matrix
partition function in terms of the complex eigenvalues $z_{j=1,\ldots,N}$ 
of a complex
$N\times N$ matrix 
$J=H + i \sqrt{\frac{1-\tau}{1+\tau}} A$ is defined as
\be
{\cal Z}_N^{(N_f,\nu)}(\tau) \equiv \int \prod_{j=1}^N
dz_j dz_j^\ast \wa(z_j)
|\Delta_N(z_1^2,\ldots,z_N^2)|^2, 
\label{Zev}
\ee
where we have introduced the weight function
\be
\wa(z)=|z|^{2a+1}\! 
\exp\left[
-\frac{N}{1-\tau^2}\left(|z|^2 -\frac{\tau}{2}(z^2+\zc^2)\right)
\right]. 
\label{weight}
\ee 
It contains $N_f$ massless quark flavors in a fixed sector of topological
charge $\nu$, where $a=N_f+\nu$. We have taken the absolute value of the 
Dirac determinant but the weight depends on the real and imaginary part of 
the eigenvalues.
The weight has been chosen to be Gaussian with variance $(1+\tau)/2N$ for both
the Hermitian and anti-Hermitian part of $J$, $H$ and $A$ respectively. 
The parameter $\tau\in[0,1]$ controls the degree of non-Hermiticity and will
be related to the chemical potential $\mu$ in QCD below.
The Jacobi determinant from the diagonalization of $J$ \cite{Gin}
yields the Vandermonde determinant  
$\Delta_N(z_1,\ldots,z_N)=\prod_{k>l}^N (z_k-z_l)$. 
The reason for the model eq. (\ref{Zev}) being 
chiral can be seen as follows.
For real eigenvalues the chiral ensemble \cite{SV}  
is usually defined on the real positive line, 
$\int_0^\infty d\lambda \lambda^a\exp{(-N\lambda)}$. By substituting 
$\lambda=y^2$ it can be mapped to the full real axis, 
$\int_{-\infty}^\infty dy\,|y|^{2a+1} \exp{(-Ny^2)}$, where $y$ corresponds 
to a real eigenvalue of the Dirac operator $\Dirac$ \cite{remark}.
In the latter picture the continuation into the complex
plane is straightforward, leading to the ensemble (\ref{Zev}). 
The only difference to the massless 
non-chiral ensemble \cite{A01} is the additional
power in $|z|$. Furthermore, the orthogonal polynomials 
for the ensemble (\ref{Zev}) are given by 
Laguerre polynomials $L_n^a(z^2)$ in the complex plane. 

A chemical potential $\mu$ 
is included in the QCD action by adding $\mu\ga_0$ to the Dirac
operator $\Dirac$ and it renders its eigenvalues complex. 
Our aim is to describe the local fluctuation of small eigenvalues close to 
the origin, because of their importance for chiral symmetry breaking 
through the Banks-Casher relation \cite{BC}. For $\mu$ not to completely 
dominate the Dirac determinant we will also restrict ourselves to small 
values of $\mu$. 

In \cite{Steph} a slightly different matrix model was introduced replacing 
$\Dirac$ by a chiral random matrix and 
keeping the additional term $\mu\ga_0$ explicitly. 
Here, we treat $\Dirac+\mu\ga_0$ as a chiral, complex matrix instead. 
The advantage of our model is that we can directly calculate all 
microscopic correlation functions. 
We will take the model \cite{Steph} to relate our parameter 
$\tau$ to $\mu$ by comparing to the
macroscopic spectral density $\rho(z)$ and its boundary 
calculated there as a function of $\mu$. 
In the limit of small $\mu$ the spectral density 
of \cite{Steph} becomes approximately constant, 
$\rho(z)=1/4\pi\mu^2$, 
and it is bounded by an ellipse, 
$x^2/4 + y^2/4\mu^2=1$, where $z=x+iy$. 
This behavior can also be observed for lattice data with small $\mu$ (e.g. in 
\cite{Bar,MPW}).
The macroscopic density in our model can be read off from \cite{So} since the
Dirac determinants are subdominant in the macroscopic large-$N$ limit:
\be
\rho(z)= \frac{1}{\pi(1-\tau^2)}\ , \ \ 
\ \mbox{if} \ \frac{x^2}{(1+\tau)^2}+\frac{y^2}{(1-\tau)^2}\leq 1 \ .
\label{rhomacro} 
\ee
We therefore identify 
\be
4\mu^2 \ \equiv\ (1-\tau^2) \ ,
\label{mutau}
\ee
valid for small chemical potential and $\tau$ close to unity meaning small
non-Hermiticity. 
For large values of $\mu$ the eigenvalue density on the lattice is no 
longer constant and develops a hole in the middle (see e.g. 
\cite{Bar,MPW}). We will also see such a hole develop in the microscopic 
correlations in the limit of strong non-Hermiticity, as shown in Fig.
\ref{strongplot} below.

After having identified all parameters in our model eq. (\ref{Zev}) we turn to
its solution using the powerful method of orthogonal polynomials \cite{Mehta}.
We only state the results and refer to \cite{GA} for
details. All eigenvalue correlation functions 
are first given for a finite number of eigenvalues $N$
and then in two different large-$N$ limits corresponding 
to weak and strong non-Hermiticity.
The orthogonal polynomials in the complex plane are defined as 
\be
\int dz dz^\ast\  \wa (z)
P_k^{(a)}(z)P_l^{(a)}(\zc) \ =\ \delta_{kl} \ .
\label{OP}
\ee
Following standard techniques \cite{Mehta} the knowledge of the 
kernel of orthogonal polynomials 
\be
K^{(a)}_N(z_1,z_2^\ast)\equiv 
[\wa(z_1)\wa(z_2^\ast)]^{\frac12}
\sum_{k=0}^{N-1} P_k^{(a)}(z_1)P_k^{(a)}(z_2^\ast) 
\label{kernel}
\ee
allows to calculate all $k$-point correlation functions 
\be
\rho_N^{(a)}(z_1,\ldots,z_k) \ =\
\det_{1\leq i,j\leq k}\left[K_N^{(a)}(z_i,z_j^\ast)\right].
\label{MM}
\ee
The result for the orthogonal polynomials eq. (\ref{OP}) is given in terms of
Laguerre polynomials
\be
P_k^{(a)}(z)\equiv \left[ 
\left( 
\begin{array}{c}
a+k\\
k
\end{array}
\right)
\fat \right]^{-\frac12}\!
(-\tau)^k L_k^a\left(\frac{Nz^2}{2\tau}\right) ,
\label{Laguerre}
\ee
with the normalization integral
\beq
&&\fat\ \equiv\ \int dzdz^\ast \  \wa (z) \label{fdef}\\
&&= N^{-a-\frac32}\pi \Gamma\left(a+\frac32\right)
(1-\tau^2)^{\frac{a}{2}+\frac34} 
P_{a+\frac32}\left(\frac{1}{\sqrt{1-\tau^2}}\right) 
\nn
\eeq
and $P_{a+\frac32}(x)$ being the Legendre function. 
All $k$-point correlation functions then follow by inserting the polynomials
into eqs. (\ref{kernel}) and (\ref{MM}). In our results the parameter 
$a=N_f+\nu$ can be kept real (with $a>-1$). For example we can set 
$a=-\frac12$ as a check, recovering the even 
subset of the Hermite polynomials in the
complex plane \cite{PdF}. 

Since we did not find eq. (\ref{Laguerre}) in the literature we briefly sketch
its derivation. Performing a change of variables $z\to \mbox{e}^{i\varphi}z$ in
the normalization integral eq. (\ref{fdef}) we obtain
\be
1=
\left\langle 
\exp\left[ \frac{u}{u-1}\left(\frac{Nz^2}{2\tau}\right)\right]
\exp\left[ \frac{\uc}{\uc-1}\left(\frac{N\zc^2}{2\tau}\right)\right]
\right\rangle ,
\label{Lrel}
\ee
with
\be
u \ \equiv\ 
\frac{\tau^2(1-\mbox{e}^{2i\varphi})}{(1-\tau^2\mbox{e}^{2i\varphi})}
\ee
and the average taken with respect to $\wa(z)$.
Multiplying both sides of eq. (\ref{Lrel}) with $((1-u)(1-\uc))^{-a-1}$
and recognizing the generating functional of the
Laguerre polynomials we obtain the desired orthogonality relation,
given properly normalized in eq. (\ref{Laguerre}).

After giving the exact solution for 
finite $N$ we turn to the large-$N$ limit. We first consider the
weak non-Hermiticity limit. 
Following \cite{FKS} we take the limit $\tau\to1$ such
that the combination 
\be
\lim_{N\to\infty}N(1-\tau^2)\ \equiv \ \al^2 \ =\ 4N\mu^2 
\label{mumicro} 
\ee
is kept fixed. Because of the identification
eq. (\ref{mutau}) we consequently also rescale $\mu$ going to zero when
$N\to\infty$.
In other words the weak non-Hermiticity parameter $\al^2$ directly measures
the chemical potential in the microscopic scaling limit. 
Such a rescaling is similar to that of the quark masses  \cite{Poulmass}. 
It has been already mentioned in \cite{HJV} that in a RMM the numerical 
effort to obtain convergence grows exponentially with 
$N\mu^2$. Keeping it fixed here should make a comparison to data feasible. 

Furthermore, we also rescale the complex eigenvalues 
keeping 
\be
N(\re\ z+i\im\ z)\ =\ Nz \ \equiv \xi \ \ \ ,
\label{microweak}
\ee
fixed. The matrix size $N$ corresponds to the volume on the lattice.
This defines our microscopic origin scaling limit in the complex
plane. The kernel eq. (\ref{kernel}) and correlators eq. (\ref{MM})
also have to be rescaled with the mean level spacing
$1/N$ of the eigenvalues. 

In order to obtain the microscopic kernel from eq. (\ref{kernel})
we replace the sum by an integral, 
$\sum_{k=0}^{N-1}\to N\int_0^1dt$, where $t=\frac{k}{N}$, and 
use the asymptotic limit of the Laguerre polynomials
to finally arrive at
\beq
K_S^{(a)}(\xi_1,\xi_2^\ast) &=& 
\frac{|\xi_1\xi_2^\ast|^{a+\frac12}}{\sqrt{2\pi\al^2}(\xi_1\xi_2^\ast)^a}
\mbox{e}^{-\frac{1}{\al^2}
\left((\Im m\xi_1)^2+(\Im m\xi_2^\ast)^2\right)}
\nn\\
&\times&
\int_0^1 dt\ \mbox{e}^{-\al^2t} 
J_a(\sqrt{2t}\xi_1)J_a(\sqrt{2t}\xi_2^\ast) .
\label{weakkernel}
\eeq
The microscopic, weakly non-Hermitian correlation functions 
obtained from eq. (\ref{MM}) are our first main
result:
\beq
&&\rho^{(a)}_S(\xi_1,\ldots,\xi_k) =\prod_{l=1}^k
\left(
\frac{|\xi_l|}{\sqrt{2\pi\al^2}}
\mbox{e}^{-\frac{2}{\al^2}(\Im m\xi_l)^2}
\right)\nn\\
&&\times
\det_{1\leq i,j\leq k}\left[
\int_0^1 dt\ \mbox{e}^{-\al^2t} 
J_a(\sqrt{2t}\xi_i)J_a(\sqrt{2t}\xi_j^\ast)
\right]\! .
\label{weakrho}
\eeq
As an important check in the Hermitian limit $\al^2\to 0$ 
corresponding to $\tau=1$ the universal 
correlations \cite{SV,ADMN} of the chiral RMM with real eigenvalues are 
reproduced. 

\begin{figure}[h]
\centerline{
\epsfig{figure=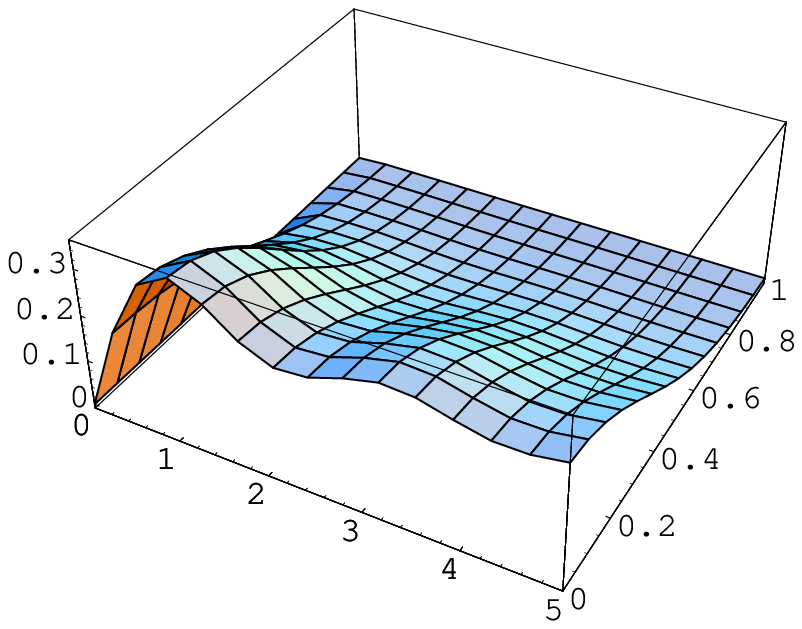,width=19pc}
\put(-160,20){$\re\ \xi$}
\put(-30,40){$\im\ \xi$}
\put(-230,140){$\rho_S^{(0)}(\xi)$}
}
\caption{
The quenched microscopic density for $\al^2=0.6$
}
\label{weakplot}
\end{figure}

To give an example for eq. (\ref{weakrho}) we have depicted 
the quenched microscopic density 
in part of the complex plane 
in Fig \ref{weakplot}. The other directions follow from symmetry. 
The oscillations known from the real case \cite{SV} spread into the complex 
plane, indicating the locations of the individual eigenvalues.

We now turn to the strong non-Hermiticity limit. In this limit $\tau\in[0,1)$ 
and consequently also $\mu$ from eq. (\ref{mutau}) 
is kept fixed in the large-$N$ limit. The eigenvalues are now rescaled with 
the square root of the volume \cite{FKS,A01}, 
\be
\sqrt{N}\,(\re\ z+i\im\ z)\ =\ \sqrt{N}\,z \ \equiv \xi \ \ \ ,
\label{microstrong}
\ee
defining our microscopic origin limit at strong non-Hermiticity. 
In this limit the infinite sum in the kernel eq. (\ref{kernel}) can be 
evaluated using standard formulas for Laguerre polynomials and we obtain
\beq
&&K_S^{(a)}(\xi_1,\xi_2^\ast)=
\frac{2^a\ \Gamma(a+1)}{N^{a+\frac32}\fat (1-\tau^2)}
\frac{|\xi_1\xi_2^\ast|^{a+\frac12}}{(\xi_1\xi_2^\ast)^a} 
\label{strongkernel}\\
&&\times 
\mbox{e}^{\frac{-1}{2(1-\tau^2)}\left(|\xi_1|^2 + |\xi_2^\ast|^2
-\frac{\tau}{2}({\xi^\ast_1}^2 -\xi_1^2 +\xi_2^2
-{\xi^\ast_2}^2)\right)}
I_a\left(\frac{\xi_1\xi_2^\ast}{1-\tau^2}\right).
\nn
\eeq
The correlation functions in the strong limit then read
\be
\rho^{(a)}_S(\xi_1,\ldots,\xi_k) \sim\prod_{l=1}^k
|\xi_l|
\ \mbox{e}^{\frac{-1}{1-\tau^2}|\xi_l|^2}
\!\det_{1\leq i,j\leq k}\left[
I_a\left(\frac{\xi_i\xi_j^\ast}{1-\tau^2}\right)
\right]
\label{strongrho}
\ee
where we have suppressed the normalization constant.
They can also be obtained from the correlators eq. (\ref{weakrho}) in the 
weak limit by taking the limit $\al\to\infty$ there and identifying 
$\al^2=1-\tau^2$. 
They differ from ref. \cite{Janco} 
due to the different 
interaction term in eq. (\ref{Zev}).
As an example for eq. (\ref{strongrho}) 
we give the quenched microscopic density in Fig. 
\ref{strongplot}.
\begin{figure}[h]
\centerline{
\epsfig{figure=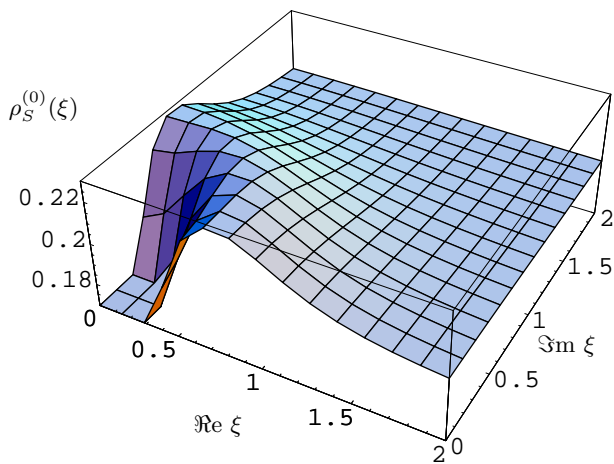,width=19pc}
\put(-160,20){$\re\ \xi$}
\put(-30,50){$\im\ \xi$}
\put(-230,140){$\rho_S^{(0)}(\xi)$}
}
\caption{
The quenched microscopic density for $\tau=0.5$
}
\label{strongplot}
\end{figure}
The microscopic spectral density develops a hole at the origin and becomes 
flat for large values. This has indeed been observed in quenched lattice data
for intermediate values of the chemical potential \cite{Bar,MPW}. At large 
$\mu$ the eigenvalues form a thin ring. 
The qualitative picture of the microscopic density in Fig. \ref{strongplot} 
remains unchanged in the presence of massless flavors $N_f$, although the 
barrier present in the figure gets flattened out.

In summary we have introduced a new class of a chiral RMM having complex 
eigenvalues. All $k$-point correlation functions have been calculated 
explicitly at finite-$N$ as well as in the limits of weak and strong 
non-Hermiticity. The parameter $\tau$
that governs the non-Hermiticity has been related to 
the chemical potential $\mu$ as it occurs in the QCD Dirac operator. 
While the microscopic density shows qualitative features of Dirac 
operator eigenvalues calculated in lattice QCD a quantitative comparison 
with data remains to be done.

Acknowledgments: 
I wish to thank P. Di Francesco, P. Forrester and E. Kanzieper for enjoyable 
discussions and correspondence.
This work was supported by the European network on ``Discrete Random 
Geometries'' HPRN-CT-1999-00161 
(EUROGRID).


\begin{thebibliography}{XX}
\bibitem{remark}In our conventions Dirac eigenvalues at $\mu=0$ are real.
\end{thebibliography}
\end{document}